\documentclass[10pt,letterpaper]{article}
\usepackage[top=0.85in,left=2.75in,footskip=0.75in]{geometry}

\usepackage{changepage}

\usepackage[utf8]{inputenc}

\usepackage{textcomp,marvosym}

\usepackage{fixltx2e}

\usepackage{amsmath,amssymb}

\usepackage{cite}

\usepackage{nameref,hyperref}

\usepackage[right]{lineno}

\usepackage{microtype}
\DisableLigatures[f]{encoding = *, family = * }

\usepackage{rotating}

\usepackage{float}
\usepackage{graphicx}
\usepackage{subfigure}

\raggedright
\usepackage[aboveskip=1pt,labelfont=bf,labelsep=period,justification=raggedright,singlelinecheck=off]{caption}

\makeatletter
\renewcommand{\@biblabel}[1]{\quad#1.}
\makeatother

\date{}

\usepackage{lastpage,fancyhdr,graphicx}
\usepackage{epstopdf}
\fancyheadoffset[L]{2.25in}
\fancyfootoffset[L]{2.25in}

\begin{document}
\vspace*{0.35in}

\begin{flushleft}
{\Large
\textbf\newline{Trend of Narratives in the Age of Misinformation}
}
\newline
\\
Alessandro Bessi\textsuperscript{1, 2\Yinyang},
Fabiana Zollo\textsuperscript{2,\Yinyang},
Michela Del Vicario\textsuperscript{2,\textcurrency a},
Antonio Scala\textsuperscript{2,3\ddag},
Guido Caldarelli\textsuperscript{2,3,4\ddag},
Walter Quattrociocchi\textsuperscript{2,*},
\\
\bigskip
\bf{1} IUSS Pavia, Italy
\\
\bf{2} IMT Lucca, Italy
\\
\bf{3} ISC-CNR, Rome, Italy
\\
\bf{4} LIMS, London, UK
\\
\bigskip

%
%





* walter.quattrociocchi@imtlucca.it

\end{flushleft}
\section*{Abstract}
Social media enabled a direct path from producer to consumer of contents changing the way users get informed, debate, and shape their worldviews. Such a {\em disintermediation} weakened consensus on social relevant issues in favor of rumors, mistrust, and fomented conspiracy thinking -- e.g.,  chem-trails inducing global warming, the link between vaccines and autism, or the New World Order conspiracy.

In this work, we study through a thorough quantitative analysis how different conspiracy topics are consumed in the Italian Facebook. 
By means of a semi-automatic topic extraction strategy, we show that the most discussed contents semantically refer to four specific categories: {\em environment}, {\em diet}, {\em health}, and {\em geopolitics}. 
We find similar patterns by comparing users activity (likes and comments) on posts belonging to different semantic categories. However, if we focus on the lifetime -- i.e., the distance in time between the first and the last comment for each user -- we notice a remarkable difference within narratives -- e.g., users polarized on geopolitics are more persistent in commenting, whereas the less persistent are those focused on diet related topics. 
Finally, we model users mobility across various topics finding that the more a user is active, the more he is likely to join all topics. Once inside a conspiracy narrative users tend to embrace the overall corpus.

\section*{Introduction}
According to \cite{furedi2006culture}, causation is bound to the way communities attempt to make sense to events and facts. 
Such a phenomenon is particularly evident on the web where users, immersed in homophile and polarized clusters, process information through a shared system of meaning \cite{bessi2014science,mocanu2014}.
Indeed, social media enabled a direct path from producers to consumers of contents -- i.e., disintermediation -- changing the way users get informed, debate, and shape their opinions \cite{brown2007word,Richard2004,QuattrociocchiCL11,Quattrociocchi2014,Kumar2010}.
However, confusion about causation encourages speculation, rumors, and mistrust \cite{sunstein2009conspiracy}.
In 2011 a blogger claimed that Global Warming was a fraud aimed at diminishing liberty and democratic tradition \cite{Forbes2011}.
More recently, rumors about Ebola caused disruption to health-care workers \cite{RumorEbola1,RumorEbola2,RumorEbola3}.
Such a scenario fostered the production of an impressive amount of conspiracy-like narratives aimed at explaining reality and its phenomena, and provide an unprecedented opportunity to study the dynamics of narratives' emergence, production, and popularity.

Recently, we pointed out that the more the users are exposed to unsubstantiated rumors, the more they are likely to jump the credulity barrier \cite{mocanu2014,bessi2014science}.  
As pointed out by \cite{kuklinski2000}, individuals can be uninformed or misinformed, and the actual means of corrections in the diffusion and formation of biased beliefs are not effective. In particular, in \cite{SOCINFO14} online debunking campaigns have been shown to create a reinforcement effect in usual consumers of conspiracy stories.
Narratives grounded on conspiracy theories have a social role in simplifying causation because they tend to reduce the complexity of reality and are able at the same time to contain the uncertainty they generate \cite{byford2011conspiracy,finerumor,hogg2011extremism}.
Conspiracy theories create a climate of disengagement from mainstream society and from officially recommended practices \cite{bauer1997resistance} -- e.g. vaccinations, diet, etc. 

Despite the enthusiastic rhetoric about the {\em collective intelligence} \cite{surowiecki2005wisdom,WelinderBBP10} the role of socio-technical systems in enforcing informed debates and their effects on the public opinion still remain unclear.
However, the World Economic Forum listed massive digital misinformation as one of the main risks for modern society \cite{Davos13}. 
A multitude of mechanisms animates the flow and acceptance of false rumors, which in turn create false beliefs that are rarely corrected once adopted by an individual \cite{Garrett2013,Meade2002,koriat2000,Ayers98}. The process of acceptance of a claim (whether documented or not) may be altered by normative social influence or by the coherence with the system of beliefs of the individual  \cite{Zhu2010,Loftus2011}.  
A large body of literature addresses the study of social dynamics on socio-technical systems from social contagion up to social reinforcement \cite{Onnela2010,Ugander2012,Lewis2012,kleinberg2013analysis,eRep,QuattrociocchiCL11,QuattrociocchiPC09,Bond2012,Centola2010,Quattrociocchi2014}.

In this work we analyze a collection of conspiracy news sources in the Italian Facebook. In particular, we focus on the emergence of conspiracy topics and the way they are consumed. We identify pages diffusing conspiracy news -- i.e. pages promoting contents {\em neglected} by main stream media. We define the space of our investigation with the help of Facebook groups very active in debunking conspiracy theses ({\em Protesi di Protesi di Complotto}, {\em Che vuol dire reale}, {\em La menzogna diventa verita e passa alla storia}). We categorize pages according to their contents and their self description.

Concerning conspiracy news pages, their self claimed mission is to inform people about topics neglected by main stream media.  
Through a semi-automatic topic extraction strategy, we show that the most discussed contents refer to four well specified topics: environment, diet, health, and geopolitics. Such topics are consumed in a very similar way by their respective audience -- i.e, users activity in terms of likes and comments on posts belonging to different categories are similar and resolves in comparable information consumption patterns. 
Conversely, if we focus on the lifetime --i.e., the distance in time between the first and the last comment for each user -- we notice a remarkable difference within topics. Users polarized on geopolitics subjects are the most persistent in commenting, whereas the less persistent users are those focused on diet narratives. 
Finally we analyze mobility of users across topics. 
Our findings show that users can jump independently from one topic to another, and such a probability increases with the user engagement. This work provides important insights about the fruition of conspiracy like rumors in online social media and more generally about the mechanisms behind misinformation diffusion.

\section*{Results and Discussion}
The analysis aims at characterizing the topical space in the conspiracy corpus of the Italian Facebook. We start our investigation by outlining the emerging topics and then we focus on the information consumption patterns. Finally we provide a data-driven model of users information consumption patterns.

Details about the mathematical and statistical tools as well as the data used in the analysis are described in Methods section.

\subsection*{Topics extraction and validation}
As a first step in our analysis we apply a semi-automatic topic extraction strategy aimed at classifying content.
 
We have $205,703$ posts ($98.62\%$ of the total corpus of conspiracy posts) containing a message -- i.e. a simple text or a description of the associated photo, video, or link. We build a Document-Term matrix ($205,703$ posts $\times$ $216,696$ terms) and take all the terms with more than $500$ occurrences ($1820$).
We then apply a supervised preprocessing in order to identify terms related to the conspiracy storytelling. Such a supervised task is performed by $20$ volunteers separately\footnote{We consider as \emph{conspiracy terms} only those terms labeled as conspiratorial by at least the 90\% of volunteers.}. The resulting set is composed by 159 terms. 

Then, we derive the co-occurrence network of conspiracy terms -- i.e., a graph where nodes are conspiracy terms, edges bond two nodes if the corresponding terms are found in the same post, and weights associated to edges indicate the number of times the two terms appear together in the corpus. Such a graph has $159$ nodes and $11,840$ edges. 

\begin{figure}[H]
	\centering
	\includegraphics[width=\textwidth, angle = 0]{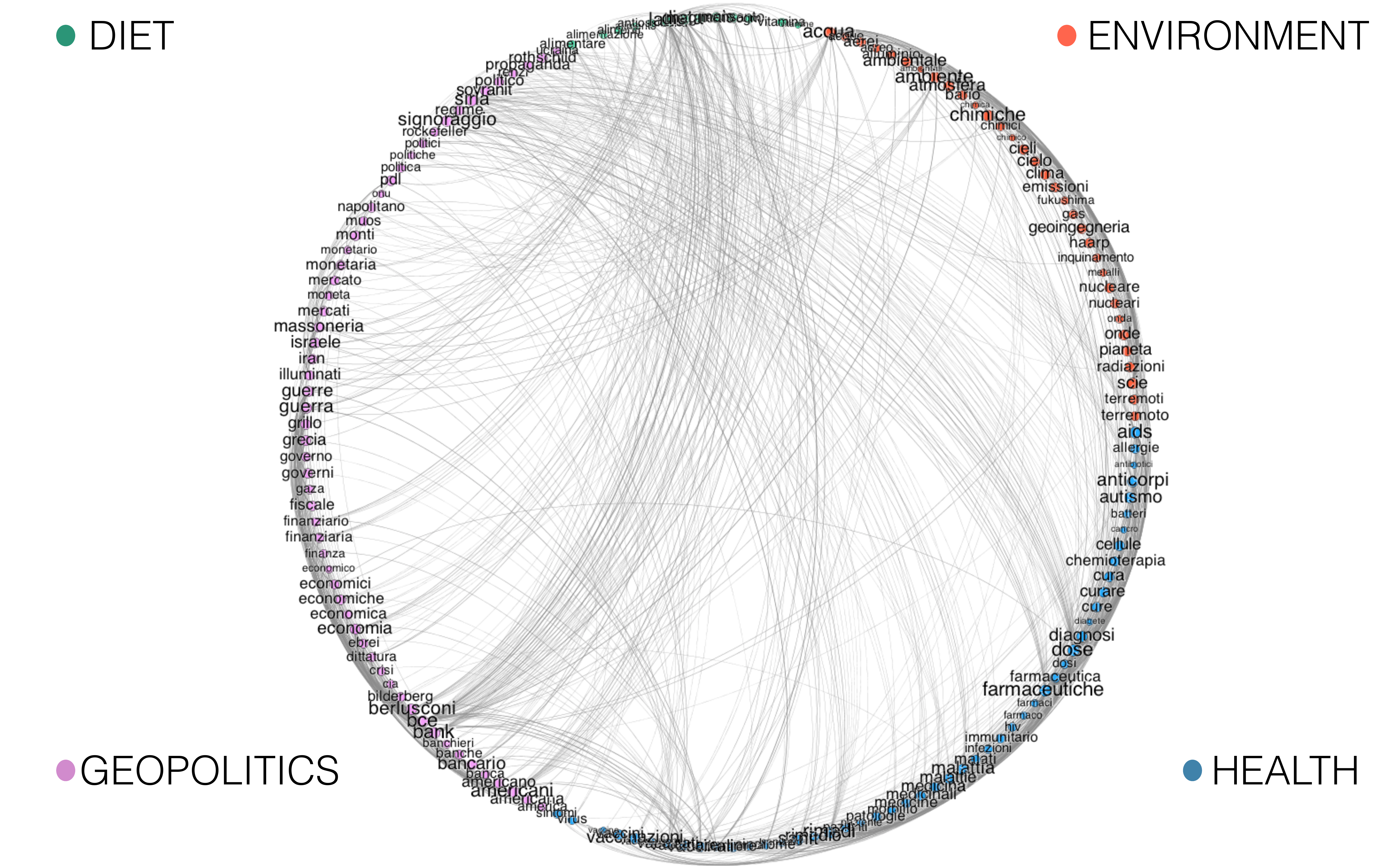} 
	\caption{\textbf{Backbone of conspiracy terms co-occurence network.} Different colors indicate nodes belonging to different conspiracy topics according to our manual classification. In particular, purple nodes belong to geopolitics, red nodes to environment, blue nodes to health, and green to diet.}
	\label{fig:manual}
\end{figure}

Since the co-occurrence network is a dense graph, we apply a filter in order to remove edges which are not significant. More precisely, we apply the disparity filter algorithm \cite{vespignani2009} (see Methods section for details) to extract the network backbone structure, thus reducing the number of edges while preserving its multi-scale nature. 
The application of the filtering algorithm with a statistical significance level of $\alpha = 0.05$ results in a graph with $159$ nodes and $1,126$ edges. We asked to the volunteers to give a generic class to each term. By accounting only for 90\% of concordance within volunteers tags on names, the semantics of conspiracy terms results in four main categories: \emph{environment}, \emph{health}, \emph{diet}, and \emph{geopolitics}. 
In Figure \ref{fig:manual} we show the backbone of the co-occurrence term network, where different colors indicate nodes belonging to different conspiracy class.

To validate the classification, we apply three different community detection algorithms -- i.e., Walktrap \cite{pons2005}, Multilevel \cite{blondel2008}, and Fast greedy \cite{clauset2004} (see Methods section for further details) -- to the backbone of conspiracy terms co-occurence network. 

\begin{figure}[H]
\centering
	\subfigure[Walktrap]
	{\includegraphics[width=0.30\textwidth]{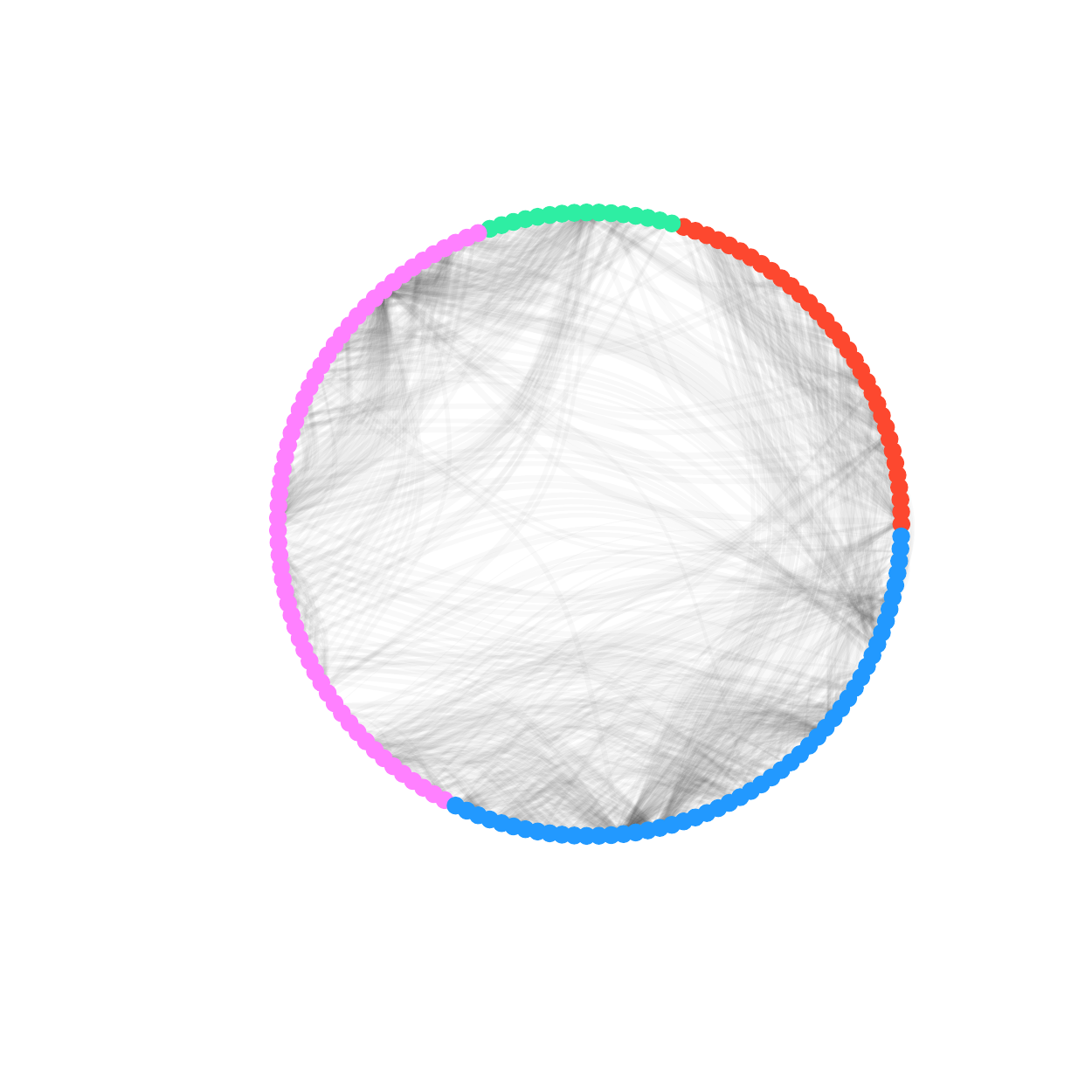}} 
	\hspace{1mm}
	\subfigure[Multilevel]
	{\includegraphics[width=0.30\textwidth]{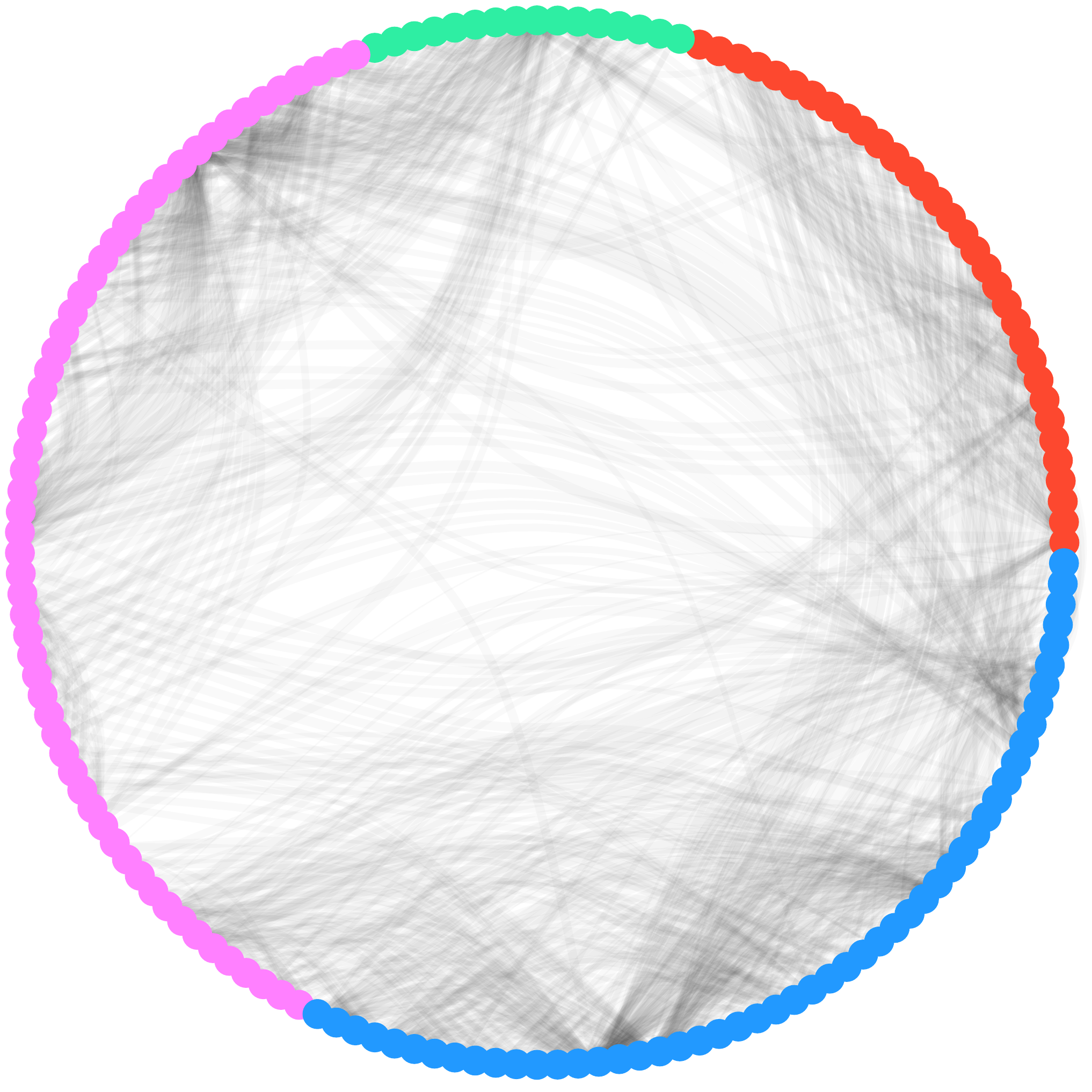}}
	\hspace{1mm}
	\subfigure[Fast greedy]
	{\includegraphics[width=0.30\textwidth]{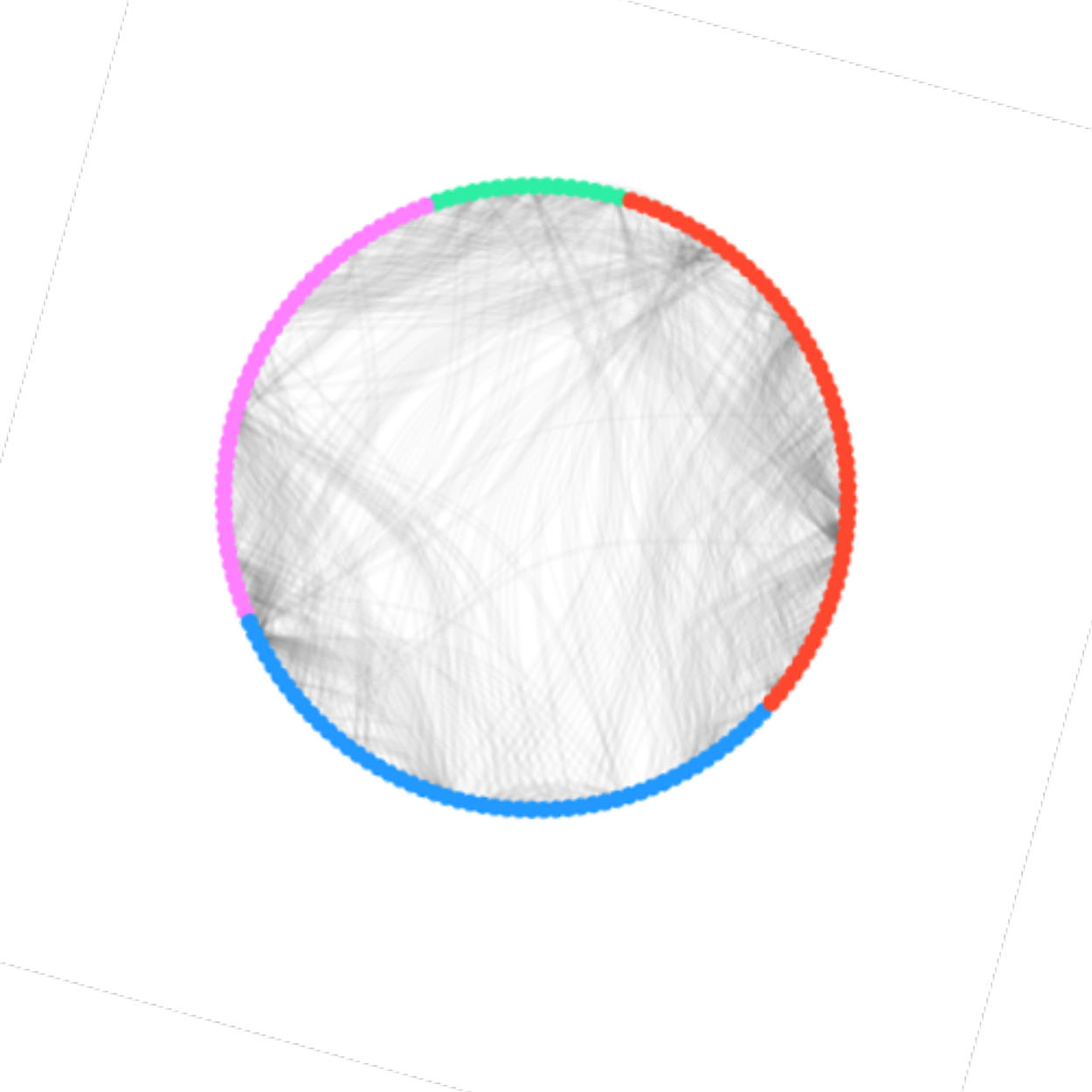}}  
\caption{\textbf{Communities of conspiracy terms}. Membership of conspiracy terms according to different community detection algorithms. Purple nodes belong to geopolitics, red nodes to environment, blue nodes to health, and green to diet.}
\label{fig:membership}
\end{figure}

In Figure \ref{fig:membership} we show the classification provided by each community detection algorithm. Multilevel and Walktrap algorithms assign each term to the same community and their accuracy with respect to our manual classification is $100\%$, while the concordance index of the Fast greedy algorithm is $88.68\%$. 

We assign a post to a given topic according to the term in it. In case of terms belonging to different topics, we apply the majority rule, in case of ties, the post is not labeled. 
Through such a criterion we are able to label $44,259$ posts -- i.e. $9,137$ environment posts, $8,668$ health posts, $3,762$ diet posts, and $22,692$ geopolitics posts.

\subsection*{Attention patterns}

\subsubsection*{Content consumption}
In order to characterize how information belonging to the different categories of conspiracy topics are consumed, we perform a quantitative analysis on users' interactions -- i.e. likes, shares, and comments\footnote{Notice that each of these actions has a particular meaning \cite{Ellison2007}. A {\em like} stands for a positive feedback to the post; a {\em share} expresses the will to increase the visibility of a given information; and a {\em comment} is the way in which online collective debates take form around the topic promoted by posts. Comments may contain negative or positive feedbacks with respect to the post.}.

\begin{figure}[H]
		\centering
		\subfigure[]
		{\includegraphics[width=0.3\textwidth]{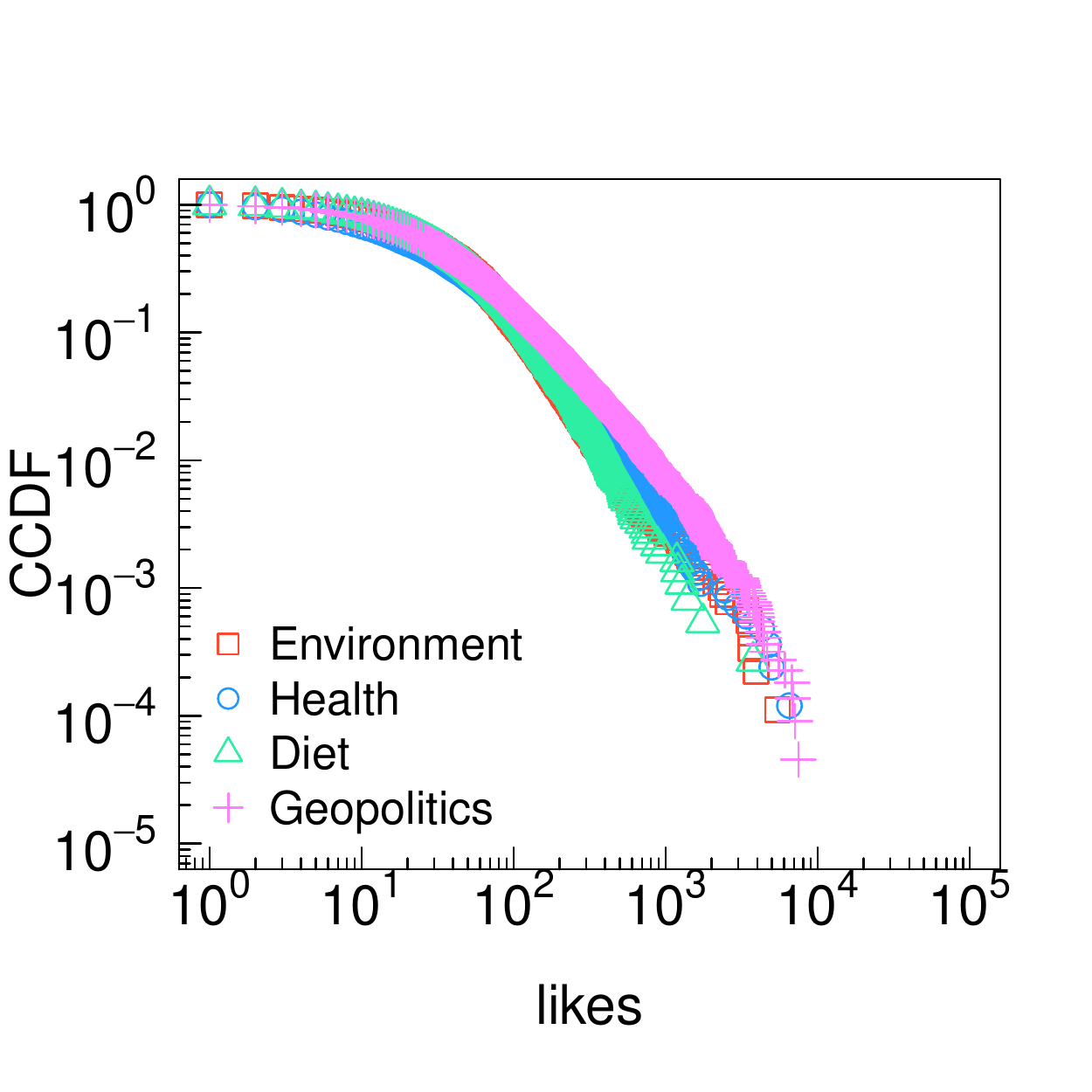}} 
		\hspace{1mm}
		\subfigure[]
		{\includegraphics[width=0.3\textwidth]{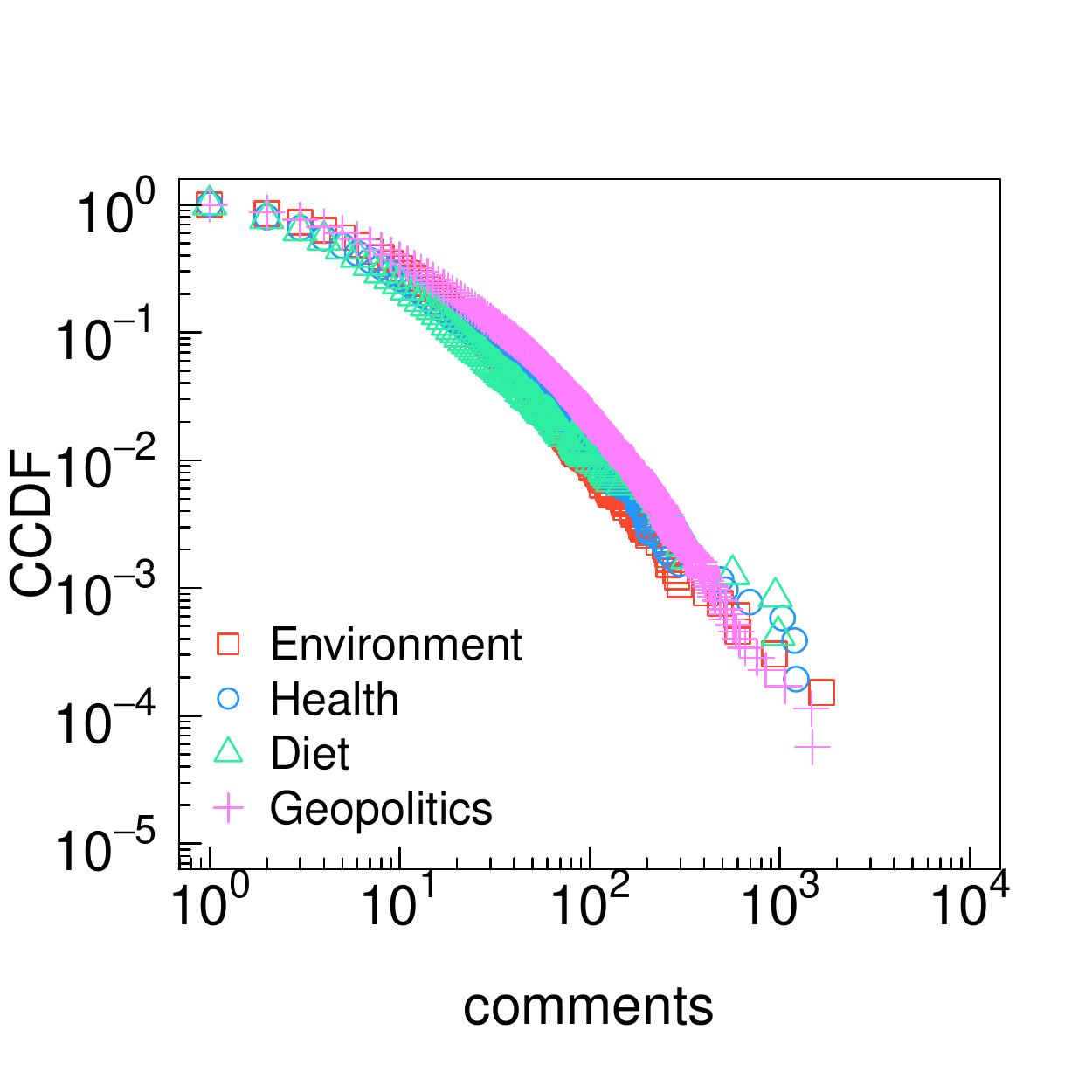}}
		\hspace{1mm}
		\subfigure[]
		{\includegraphics[width=0.3\textwidth]{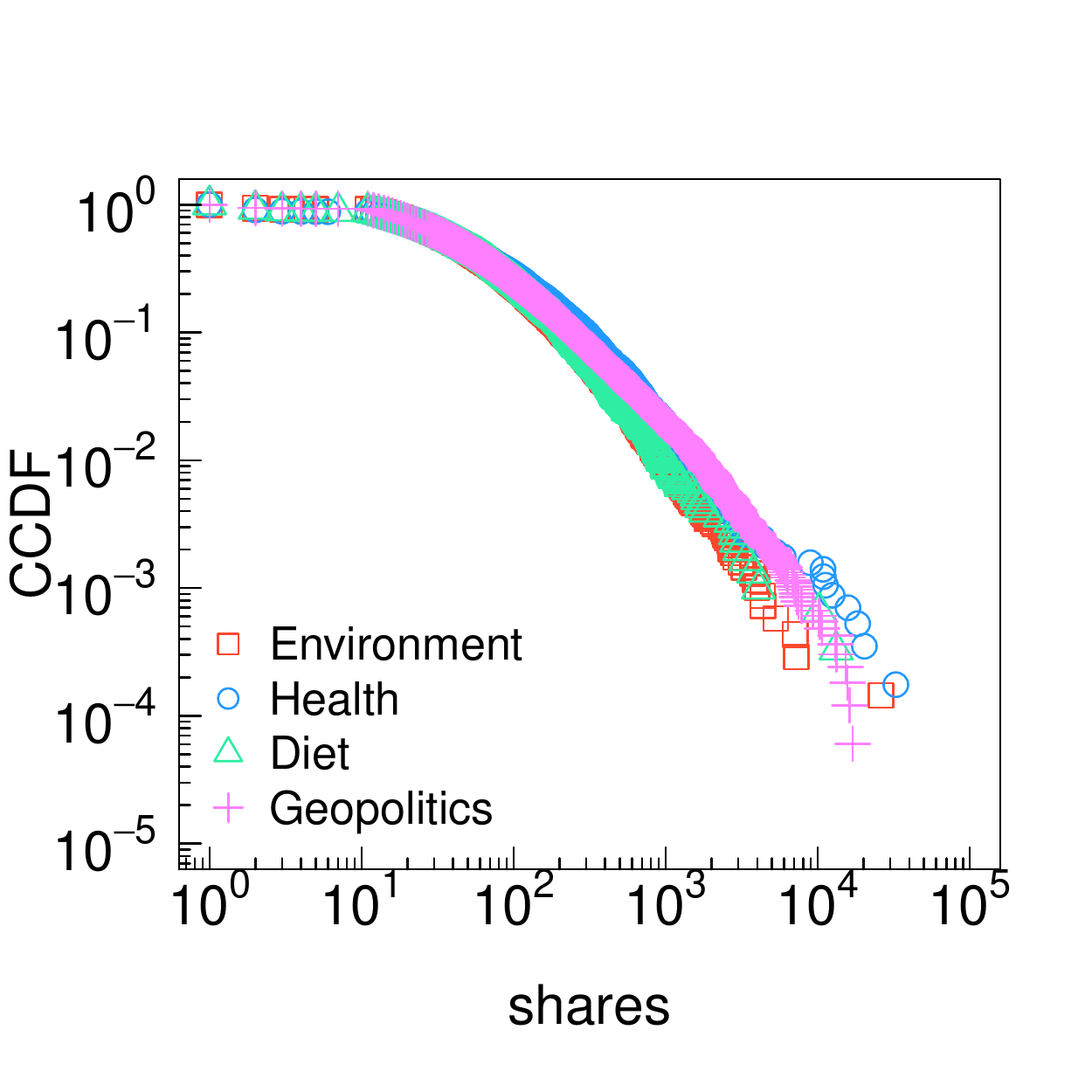}} 
		
\caption{\textbf{Topics attention patterns.} Complementary cumulative distribution functions (CCDFs) of the number of likes (a), comments (b), and shares (c) received by posts belonging to different conspiracy topics.}
\label{fig:attention_topics}
\end{figure}

In Figure \ref{fig:attention_topics} we show the complementary cumulative distribution functions (CCDFs)\footnote{We prefer to show CCDFs rather than CDFs since the former allow to plot heavy-tailed distributions in doubly logarithmic axes, and thus emphasize their long tail behavior.} of the number of likes, comments, and shares received by posts belonging to different conspiracy categories of topics. 
Likes, comments, and shares of posts are long-tailed distributed and best fitted by discrete power law distributions. 
To further characterize differences within the distributions, in Table \ref{tab:estimates_topics} we summarize the estimated lower bounds and scaling parameters for each distribution.

\begin{table}[ht]
\centering
\begin{tabular}{rcc|cc|cc}
	& \multicolumn{2}{c}{\textbf{Likes}} & \multicolumn{2}{c}{\textbf{Comments}} & \multicolumn{2}{c}{\textbf{Shares}}  \\ 
	& $\hat{x}_{min}$ & $\hat{\alpha}$ & $\hat{x}_{min}$ & $\hat{\alpha}$ & $\hat{x}_{min}$ & $\hat{\alpha}$ \\
	
	\textbf{Environment} & $142$ & $2.82$ & $42$  & $2.82$ & $408$ & $2.62$ \\ 
	\textbf{Health}      & $172$ & $2.68$ & $37$  & $2.59$ & $435$ & $2.39$ \\ 
	\textbf{Diet}        & $135$ & $2.84$ & $15$  & $2.36$ & $358$ & $2.59$ \\ 
	\textbf{Geopolitics} & $167$ & $2.36$ & $135$ & $3.14$ & $407$ & $2.25$ \\ 
	\multicolumn{7}{c}{}\\
\end{tabular}

\caption{\textbf{Power law fit of conspiracy topics attention patterns.} Lower bounds and scaling parameters estimates for the distributions of the number of likes, comments, and shares received by posts belonging to different conspiracy topics.}
\label{tab:estimates_topics}
\end{table}	

\begin{figure}[H]
	\centering
	\includegraphics[width=0.7\textwidth]{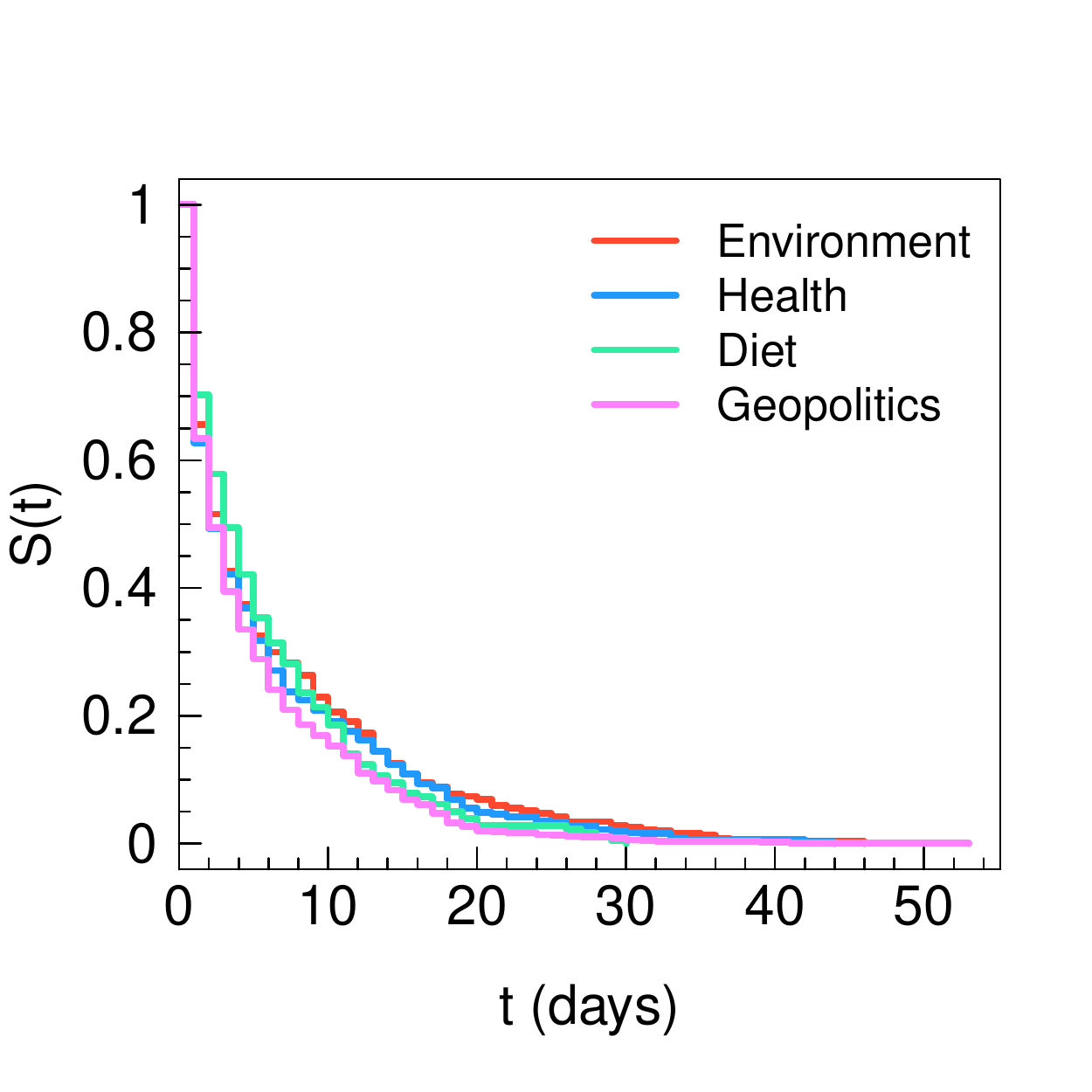} 
	\caption{\textbf{Lifetime of conspiracy topics.} Kaplan-Meier estimates of survival functions of posts belonging to different conspiracy topics.}
	\label{fig:lf_topics}
\end{figure}

To analyze lifetime of posts from different categories, we compute the temporal distance between the first and last comment for each post. In Figure \ref{fig:lf_topics} we show the Kaplan-Meier estimates of survival functions (see Methods for further details) for posts belonging to different conspiracy topics. 
The p-value associated to the Gehan-Wilcoxon test (a modification of the log-rank test) is $p = 0.091$, which lets us conclude that there are not significant statistical differences between the survival functions of posts belonging to different conspiracy topics.  

Our findings show that conspiracy topics are consumed in a similar way. In particular, we find that survival functions of posts belonging to different conspiracy topics do not show different statistical signatures.

\subsubsection*{Users activity}
Here, we consider users' attention patterns with respect to different conspiracy topics by analyzing the number of likes and comments, as well as the lifetime of each user -- i.e. the temporal distance between his first comment and last comment on a post belonging to a specific category -- that can be intended as an approximation of users persistence in online collective debating.

We consider as conspiracy users those whose liking activity on conspiracy pages is greater than the $95\%$ of their total liking activity -- i.e. a conspiracy user left at most $5\%$ of her likes on posts belonging to science pages. Such a method allows to identify $790,899$ conspiracy users. Moreover, we consider a user polarized towards a given conspiracy topic if she has more than the $95\%$ of her likes on posts belonging to that topic. Such a criterion allows to classify $232,505$ users ($29.39\%$ of the total). Table \ref{tab:users} summarizes the classification task's results. We observe that the majority of users is concerned about conspiracy stories related to geopolitics ($62.95\%$), whereas conspiracy narratives about environment ($18.39\%$) and health ($12.73\%$) attract a smaller yet substantial number of users, while diet ($5.94\%$) seems to be considered a less attractive subject. 
However, notice that here we are analyzing conspiracy users that have the most of their total activity on a given topic. Indeed, Figure \ref{fig:attention_topics} shows that diet receives as much attention as the other conspiracy topics, even if it is not a polarizing topic, as shown in Table \ref{tab:users}.

\begin{table}[h]
\centering
\begin{tabular}{rcc}
	& \textbf{Users} & \textbf{\%} \\ 
	\textbf{Geopolitics} &  $146,359$ & $62.95$ \\ 
	\textbf{Environment} & $42,750$ & $18.39$ \\ 
	\textbf{Health} & $29,587$  & $12.73$  \\ 
	\textbf{Diet} & $13,807$ &  $5.94$ \\ 
	\multicolumn{3}{c}{}
\end{tabular}
	
\caption{\textbf{Polarization of users towards different conspiracy topics.}}
\label{tab:users}
\end{table}

\begin{figure}[h]
	\centering
	\subfigure[]
	{\includegraphics[width=0.48\textwidth]{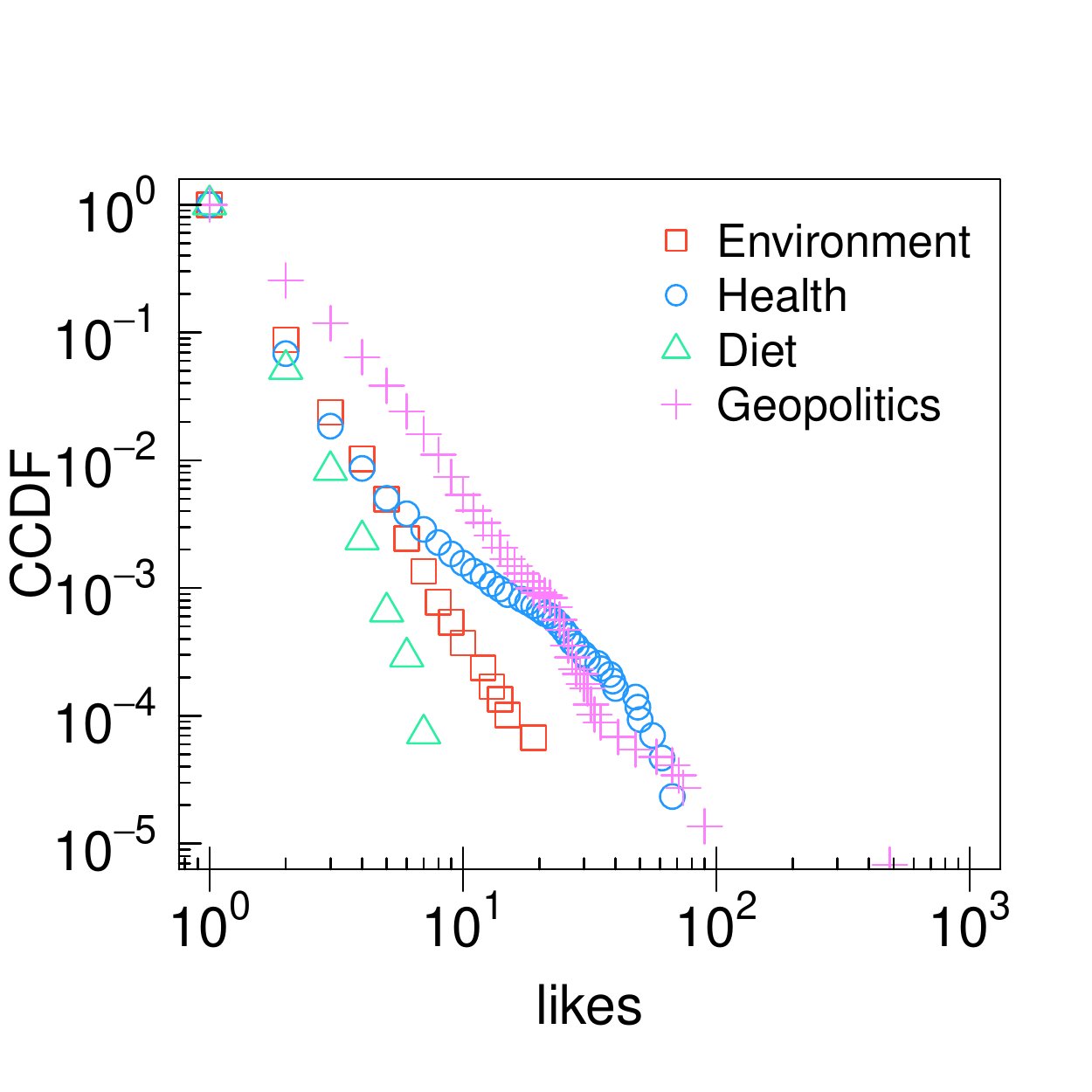}} 
	\hspace{1mm}
	\subfigure[]
	{\includegraphics[width=0.48\textwidth]{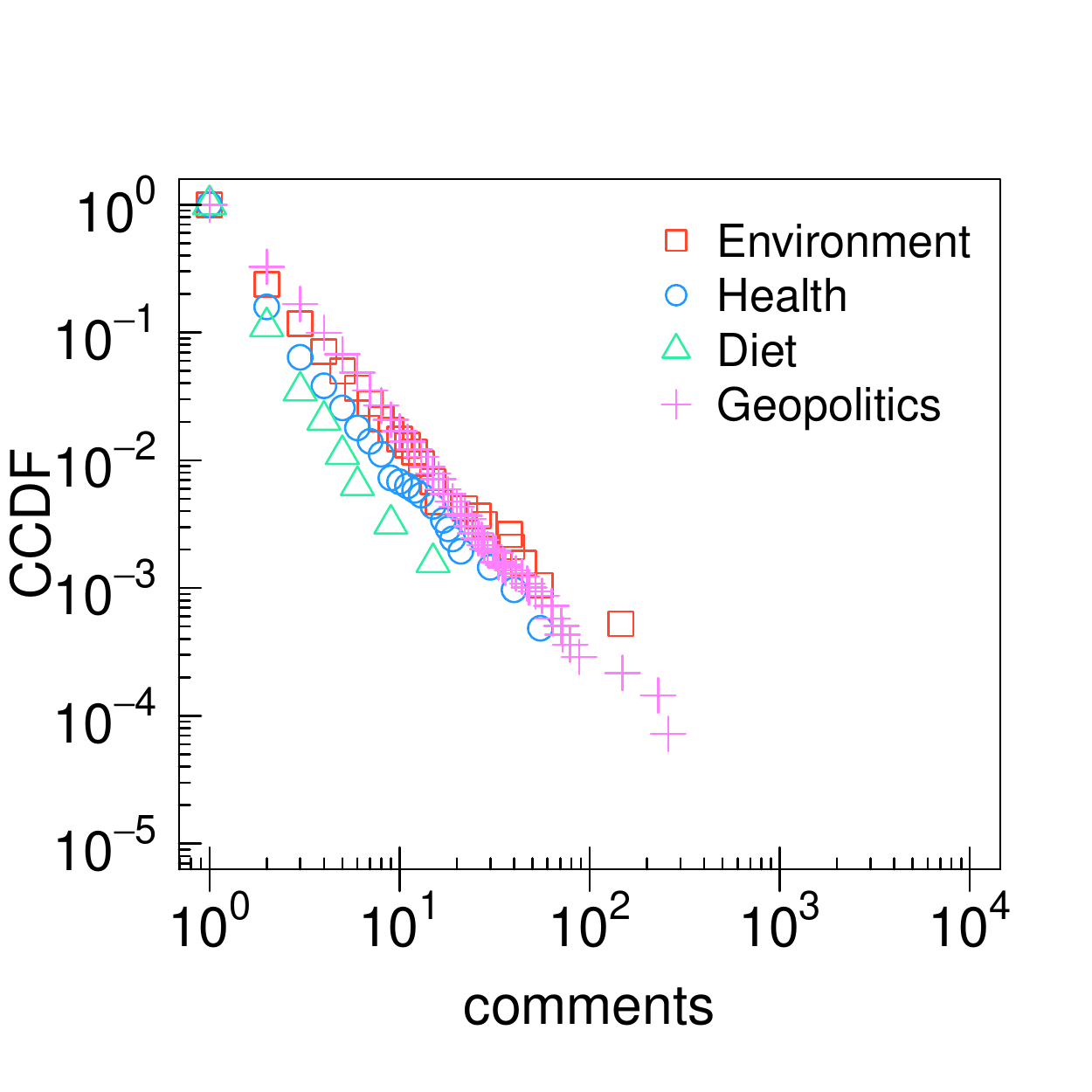}} 
	\caption{\textbf{Users attention patterns}. CCDFs of the number of likes (a) and comments (b) by users polarized on different conspiracy topics.}\label{fig:attention_users}
\end{figure}

In Figure \ref{fig:attention_users} we show the CCDFs of the number of likes and comments of users polarized towards different conspiracy topics. We observe minor yet significant differences between attention patterns of different conspiracy users. 
Table \ref{tab:estimates_users} summarizes the estimated lower bounds and scaling parameters for each distribution.
These results show that users polarized towards different conspiracy topics consume information in a comparable way -- i.e, with some differences all are well described by a power law.

\begin{table}[h]
 \centering
 \begin{tabular}{rcc|cc}
	& \multicolumn{2}{c}{\textbf{Likes}} & \multicolumn{2}{c}{\textbf{Comments}}  \\ 
	& $\hat{x}_{min}$ & $\hat{\alpha}$ & $\hat{x}_{min}$ & $\hat{\alpha}$  \\
	
	\textbf{Environment} & $5$ & $4.37$ & $3$  & $2.49$  \\ 
	\textbf{Health}      & $5$ & $2.51$ & $3$  & $2.56$  \\ 
	\textbf{Diet}        & $4$ & $5.52$ & $3$  & $2.94$  \\ 
	\textbf{Geopolitics} & $6$ & $3.61$ & $6$  & $2.88$  \\ 
	\multicolumn{5}{c}{}
\end{tabular}
\caption{\textbf{Powerlaw fit of conspiracy users attention patterns.} Lower bounds and scaling parameters estimates for the distributions of the number of likes and comments left by users polarized towards different conspiracy topics.}
\label{tab:estimates_users}
\end{table}

\begin{figure}[H]
\centering
\includegraphics[width=0.7\textwidth]{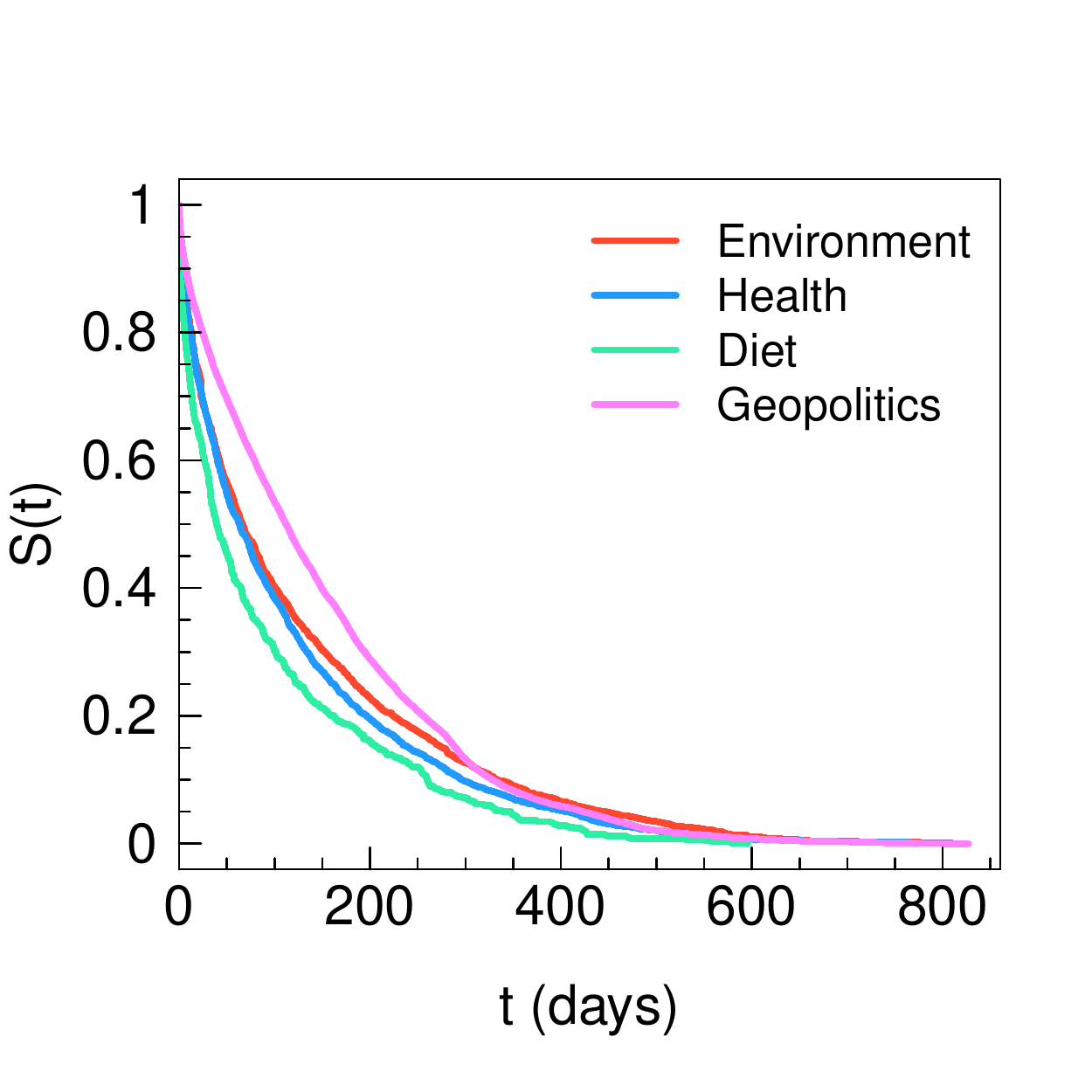} 
\caption{\textbf{Persistence of conspiracy users.} Kaplan-Meier estimates of survival functions for users polarized towards different conspiracy topics.}
\label{fig:lf_users}
\end{figure}

In order to analyze the persistence of users polarized towards different conspiracy topics, we compute the temporal distance between the first and last comment of each user on posts belonging to the conspiracy topic on which the user is polarized. 
In Figure \ref{fig:lf_users} we show the Kaplan-Meier estimates of survival functions (see Methods section for further details) for conspiracy users polarized towards different topics. 

The Gehan-Wilcoxon test assesses a significant difference between the four survival functions (all p-values are less than $10^{-6}$).

Summarizing, we observe minor yet significant differences in the way users polarized towards different conspiracy topics consume information. 
Moreover, by focusing on the lifetime -- i.e. the temporal distance between users' first and last comment -- we find a remarkable difference within those users. In particular, we notice that users polarized on geopolitics subjects are the most persistent in commenting, whereas the less persistent users are those focused on diet narratives. 

\subsection*{Modeling user mobility}
In this section we focus on users' activity across different conspiracy topics.
In Table \ref{tab:mobility} we summarize users' behavior by showing the Pearson correlations of their liking activity within the different categories of topics. We see meaningful correlations between the liking activity of users across different conspiracy topics.

\begin{table}[ht]
\centering
\begin{tabular}{rcccc}
		& \textbf{Envir} & 	\textbf{Health} & 	\textbf{Diet} & 	\textbf{GeoPol} \\ 
		\textbf{Envir} 	& 1.00 & 0.68 & 0.61 & 0.64 \\ 
		\textbf{Health} &      & 1.00 & 0.78 & 0.65 \\ 
		\textbf{Diet} 	&      &      & 1.00 & 0.48 \\ 
		\textbf{GeoPol} &      &      &      & 1.00 \\ 
		\multicolumn{5}{c}{}
\end{tabular}
\caption{\textbf{Mobility of users across topics.} Pearson correlations coefficients of conspiracy users' liking activity between different categories.}
\label{tab:mobility}
\end{table}

\begin{figure}[h]
	\centering
	\includegraphics[width = 0.7\textwidth]{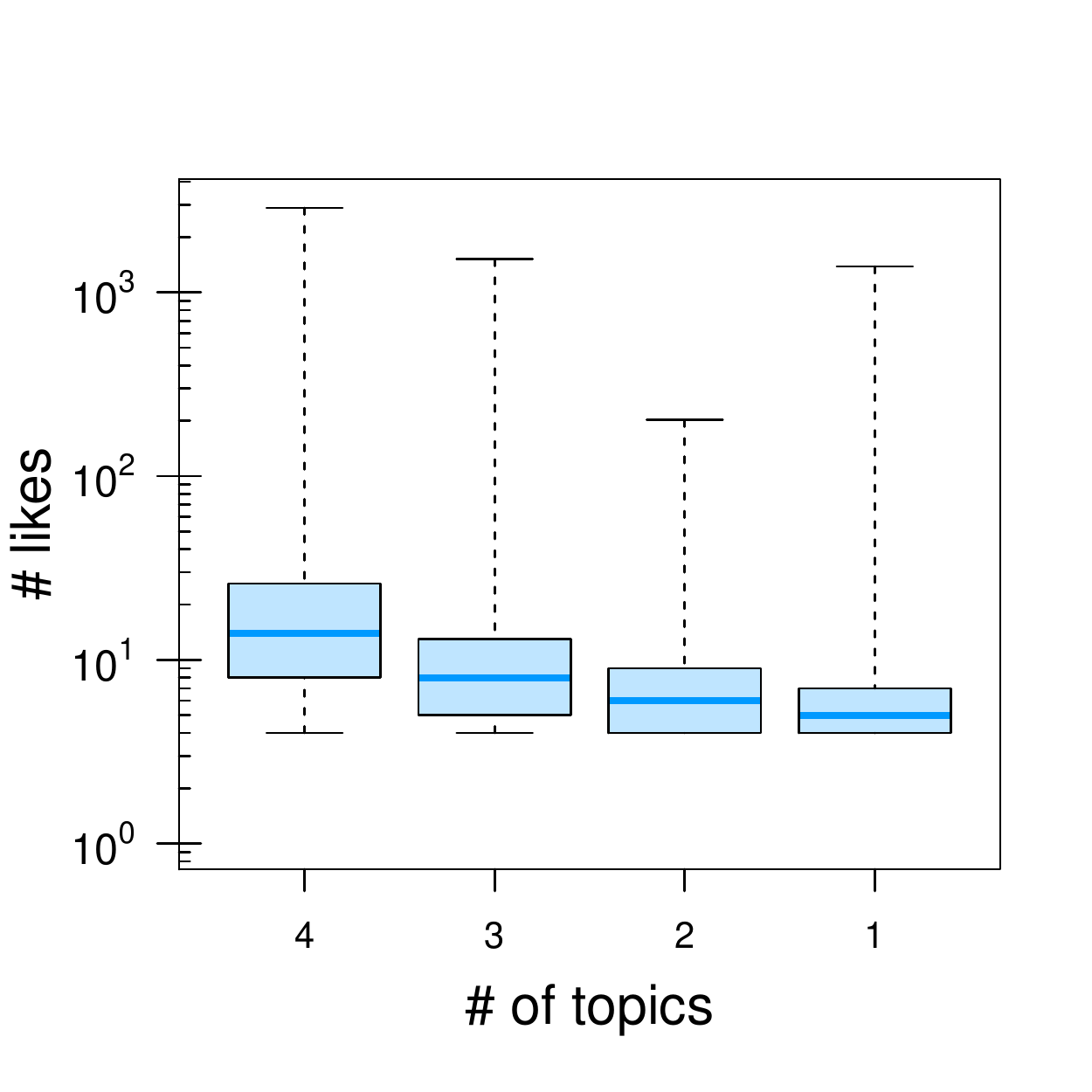}
	\caption{\textbf{Engagement and mobility across topics.} Light blue lines represent the median of the likes distributions; pale blue shaded boxes represent the interquartile range ($25$--$75$ percentile); horizontal bars represent the extremes of the distributions. Users active on $4$ topics are $15,510$; users active on $3$ topics are $20,929$; users active on $2$ topics are $21,631$; and  users active on $1$ topic are $9,980$.}
	\label{fig:boxplot}
\end{figure}

We analyze the relationship between the engagement of a user -- i.e. the number of likes she left on conspiracy posts -- and how her activity is distributed across categories. Figure \ref{fig:boxplot} shows that the more a conspiracy user is engaged the more his activity is distributed across different conspiracy topics.

By considering only users with at least $4$ likes ($n = 68,050$) -- necessary condition to be active on the four identified topics -- we can model the relationship between the number of likes and the number of considered topics by means of a proportional odds model (see Methods section for an extended discussion). 

In particular, we consider the number of topics liked by users as the ordinal dependent variable, i.e. we have $j = (K - 1) = 3$ ordered categories: $1|2$, $2|3$, and $3|4$. We consider the number of likes left by users as the predictor of our model. Thus, we need to estimate three intercepts and one regression coefficient. Table \ref{tab:prop} reports details about the performed regression.

\begin{table}[h]
\centering
\begin{tabular}{rcccc}
	\textbf{Coefficients} & \textbf{Value} & \textbf{Std. Error} & \textbf{t-value} & \textbf{p-value}\\ 
	\# of likes & 0.1141 & 0.001016 & 112.3 & $< 10^{-6}$ \\ 
	& & & & \\
	\textbf{Intercepts} & \textbf{Value} & \textbf{Std. Error} & \textbf{t-value} & \textbf{p-value}\\

    1$|$2 & -0.7602 & 0.0135 & -56.4267 & $< 10^{-6}$ \\ 
	2$|$3 & 1.0783 & 0.0126 & 85.7607 & $< 10^{-6}$ \\ 
    3$|$4 & 2.9648 & 0.0177 & 167.4990 & $< 10^{-6}$ \\ 
    \multicolumn{5}{c}{}
\end{tabular}

\caption{\textbf{Proportional Odds Model.} Log-odds regression coefficient and intercepts with related standard errors, t-values, and p-values. Confidence interval at $95\%$ for the estimated coefficient is $(0.1121, 0.1161)$. Chi-Square test's p-value is $1$, so we do not reject the null hypothesis ($H_{0}$ : current model is good enough) and conclude that the model is a good fit.}
\label{tab:prop}
\end{table}
	
The estimated coefficient, $\beta$, can be difficult to interpret because it is scaled in terms of logs. Another way to interpret these kind of regression models is to convert the coefficient into a odds ratio (see Methods section for further details). To get the odds ratio (OR) we exponentiate the estimate:
$$\mathnormal{OR} = \exp(\beta) = \exp(0.1141) = 1.12 $$

Since $\mathnormal{OR} > 1$, an increase in the number of likes left by a user raises her probability to consider a greater number of topics. In particular, an increase in the number of likes is associated with $12\%$ times increased odds of considering a higher number of topics.	 

The model provides four probabilities for each user to belong to one of the four categories. For each user we consider the category associated with the higher probability as the predicted category. In order to evaluate the goodness of fit of the model, we compare the predicted categories vs the real ones by means of the absolute distance coefficient and we find that:
$$ \delta = 1 - \frac{\sum{|d_{i}|}}{n(K-1)} = 0.852,$$

where $|d_{i}| = |x_{i} - y_{i}|$ is the absolute distance between the real and the predicted categories, $n$ is the total number of users, and $K$ is the number of categories. Since the absolute distance coefficient is close to $1$, the proposed model provides a good fit for the data.  

Summarizing, the more a user is engaged in conspiracy storytelling the more her probability to consider a higher number of different conspiracy topics. Indeed, we deliver a data-driven model of information consumption pointing out that users engagement on different topics is mainly driven by their overall commitment on conspiracy storytelling and that with the increasing of the engagement they tend to embrace the overall corpus without a specific pattern.

\section*{Conclusions}
Conspiracy theories are considered to belong to false beliefs overlooking the pervasive unintended consequences of political and social action \cite{sunstein2009conspiracy,locke2009conspiracy}.
Social media fostered the production of an impressive amount of rumors, mistrust, and conspiracy-like narratives aimed at explaining (and oversimplifying) reality and its phenomena.
Such a scenario provides an unprecedented opportunity to study the dynamics of topics emergence, production, and popularity.
Indeed, in this work we focus on how conspiracy topics are consumed in the Italian Facebook.
 
The understanding of consumption patterns behind unsubstantiated claims might provide important insight both at the level of popularity of topics as well as to prevent misinformation spreading.
Users activity in terms of likes and comments on posts belonging to different categories are similar and resolves in similar information consumption patterns. Conversely, if we focus on the lifetime -- i.e., the distance in time between the first and the last comment for each user -- we notice a remarkable difference within topics. Users polarized on geopolitics subjects are the most persistent in commenting, whereas the less persistent users are those focused on diet narratives. 
Finally we focus on the mobility of users across topics. In particular we address the patterns behind the consumption of different topics. Users can jump independently from one topic to another, and such a probability increases with the user engagement (number of likes on a specific topic). Each new like on the same category increases of the 12\% the probability to pass to a new one. Our work provides important insights about the diffusion and emergence of misinformed narratives.

\newpage
\section*{Acknowledgments}
Funding for this work was provided by EU FET project MULTIPLEX nr. 317532 and SIMPOL nr. 610704. The funders had no role in study design, data collection and analysis, decision to publish, or preparation of the manuscript.

\section*{Methods}
\subsection*{Data Collection}
We define the space of our investigation with the help of some
Facebook groups very active in the debunking of conspiracy theses. The resulting dataset is composed by $39$ public Italian Facebook pages.

Notice that the dataset is the same used in \cite{bessi2014science} and \cite{bessi2014social}. However, in this paper we focus on the 39 (exhaustive set) conspiracy pages with the intent to further characterize the attention dynamics driving the diffusion of conspiracy topics on the Italian Facebook. 
We download all posts from these pages in a timespan of 4 years (2010 to 2014). In addition, we collect all the likes and comments from the posts, and we count the number of shares. In total, we collect around $9M$ likes and $1M$ comments, performed by about $1.2M$ and $280K$ Facebook users, respectively. In Table~\ref{tab:data_dim} we summarize the details of the data collection.

\begin{center}
\begin{table}[h]
\centering
	\begin{tabular}{l|c|}
		\hline\bf { Entity  }  & \bf {Total} \\ 
		\hline 
		Pages & $ 39 $ \\
		Posts & $ 208,591 $ \\
		Likes & $ 6,659,382$  \\
		Comments & $ 836,591 $\\
		Shares &  $16,326,731$\\
		Likers & $ 864,047 $\\
		Commenters & $ 226,534 $\\
\end{tabular}\newline
\caption{ \textbf{Breakdown of the Facebook dataset.}}
\label{tab:data_dim}
\end{table}
\end{center}


\subsection*{Preliminaries and Definitions}
\paragraph{Bipartite Network.}  
We consider a bipartite network whose nodes are conspiracy posts and conspiracy terms. The presence of a term on a given post determines a link between the term and the post. More formally, a bipartite graph is a triple $\mathcal{G}=(A,B,E)$ where $A=\left\{ a_{i}\,|\,i=1,\dots, n_{A}\right\} $ and $B=\left\{ b_{j}\,|\,j=1,\dots, n_{B}\right\} $ are two disjoint sets of nodes, and $E\subseteq A\times B$ is the set of edges -- i.e. edges exist only between nodes of the two different sets $A$ and $B$. The bipartite graph $\mathcal{G}$ is described by the matrix $M$
defined as
\[
M_{ij}=\left\{ \begin{array}{cc}
1 & if\, an\, edge\, exists\, between\, a_{i}\, and\, b_{j}\\
0 & otherwise
\end{array}\right.
\]

Referring to $A$ as the set of conspiracy terms, in our analysis we use the co-occurrence matrix $C^{A}=MM^{T}$, that is the weighted adjacency matrix of the co-occurrence of conspiracy terms on conspiracy posts. Each non-zero element of $C^{A}$ corresponds to an edge among nodes $a_{i}$ and $a_{j}$ with weight $P_{ij}^{A}$.

\paragraph{Disparity Filter.} 
Disparity filter is a network reduction algorithm which extracts the backbone structure of a weighted network, thus reducing its size without destroying its multi-scale nature. In particular, the method introduced in \cite{vespignani2009} is based on the null hypothesis that the normalized weights corresponding to the connections of a given node with degree $k$ are uniformly distributed. The disparity filter identifies which links for each node should be preserved in the network. The null model allows such a discrimination through the computation -- for each edge of a given node -- of the probability $\alpha_{ij}$ that its normalized weight $p_{ij}$ is compatible with the null hypothesis. All the links with $\alpha_{ij}$ smaller than a certain significance level $\alpha$ reject the null hypothesis, and can be considered as significant heterogeneities characterizing the network. The statistically significant edges will be those whose weights satisfy the relation
$$ \alpha_{ij} = 1 - (k - 1) \int_{0}^{p_{ij}} (1-x)^{k-2}dx < \alpha, $$

indicating that by decreasing the significance level $\alpha$ we can filter out additional links, and thus focus on more relevant edges.

\paragraph{Community Detection Algorithms.} 
In order to validate our manual classification of conspiracy terms, we apply three well known community detection algorithms to the backbone of the conspiracy terms co-occurrence network.

Walktrap \cite{pons2005} computes a measure of similarities between nodes based on random walks which has several important advantages: it captures well the community structure in a network, it can be computed efficiently, and it can be used in an agglomerative algorithm to compute efficiently the community structure of a network. Such an algorithm runs in time $\mathcal{O}(mn^2)$ and space $\mathcal{O}(n^2)$ in the worst case, and in time $\mathcal{O}(n^2\log n)$ and space $\mathcal{O}(n^2)$ in most real-world cases, where $n$ and $m$ are respectively the number of nodes and edges in the network.

Multilevel \cite{blondel2008} is based on multilevel modularity optimization. Initially, each node is assigned to a community on its own. In every step, nodes are re-assigned to communities in
a local, greedy way. Nodes are moved to the community in which they achieve the highest modularity. Such an algorithm runs in linear time when $m \sim n$, where $n$ and $m$ are respectively the number of nodes and edges in the network.

Fast greedy \cite{clauset2004} it is a hierarchical agglomeration algorithm for detecting community structure. Its running time on a network with $n$ nodes and $m$ edges is $\mathcal{O}(m d \log n)$ where $d$ is the depth of the dendrogram describing the community structure. Many real-world networks are sparse and hierarchical, with $m \sim n$ and $d \sim \log n$, in which case such an algorithm runs in essentially linear time, $\mathcal{O}(n \log^2 n)$.

\paragraph{Kaplan-Meier estimator.}
Let us define a random variable $T$ on the interval $[0,\infty)$,
indicating the time an event takes place. The cumulative distribution function (CDF), $F(t) = \textbf{Pr}(T \leq t)$,
indicates the probability that such an event takes place within a given time~$t$. The survival function, defined as the complementary CDF (CCDF\footnote{We remind
that the CCDF of a random variable $X$ is one minus the CDF, the function $f(x)=\textbf{Pr}(X>x)$.}) of $T$, represents the probability that an event lasts beyond a given time period $t$. To estimate this probability we use the \emph{Kaplan--Meier estimator}~\cite{KM58}. 

Let $n_{t}$ denote the number of that commented before a given time
step $t$, and let $d_{t}$ denote the number of users that stop commenting precisely
at~$t$. Then, the estimated survival probability after time $t$ is
defined as $(n_{t} - d_{t})/n_{t}$. 
Thus, if we have $N$ observations at times $t_1\le t_2\le\dots\le t_N$,
assuming that the events at times $t_i$ are jointly independent, the Kaplan-Meier
estimate of the survival function at time $t$ is defined as
$$\hat{S}(t) = \prod_{t_{i}<t}( \frac{n_{t_i} - d_{t_i}}{n_{t_i}}).$$

\paragraph{Odds ratio.}
Probability and odds are two basic statistical terms to describe the likeliness that an event will occur. Probability is defined\footnote{For the sake of brevity we consider only the frequentist approach.} as the fraction of desired outcomes in the context of every possible outcome with a value in $[0,1]$, where 0 would be an impossible event and 1 would represent an inevitable event. Conversely, odds can assume any value in $[0,\infty)$, and they represent a ratio of desired outcomes versus undesired outcomes. Given a desired outcome $A$, the relationship between the probability $P(A)$ that event $A$ will occur, and its odds $O(A)$ is
$$ P(A) = \frac{O(A)}{1 + O(A)} \quad \mathnormal{and}\quad O(A) = \frac{P(A)}{1 - P(A)}.$$
It follows that the odds ratio (OR) of two events $A$ and $B$ is defined as
$$ OR(A,B) = \frac{O(A)}{O(B)} = \frac{\frac{P(A)}{1 - P(A)}}{\frac{P(B)}{1 - P(B)}} = \frac{P(A) [1 - P(B)]}{P(B) [1 - P(A)]}.$$

\paragraph{Proportional Odds Model.} 
The proportional odds model is a class of generalized linear models used for modeling the dependence of an ordinal response on discrete or continuous covariates.

Formally, let $Y$ denote the response category in the range $1,\dots,K$ with $K \geq 2$, and let $\pi_{j} = \mathbf{Pr}(Y \leq j\,|\,x)$ be the cumulative response probability when the covariate assumes value $x$. The most general form of linear logistic model for the $j$th cumulative response probability,
$$\mathnormal{logit}(\pi_{j}) = \mathnormal{ln}\left(\frac{\pi_{j}}{1 - \pi_{j}}\right) = \alpha_{j} + \beta^{T}_{j}x,  $$
is one in which both the intercept $\alpha$ and the regression coefficient $\beta$ depend on the category $j$. The proportional odds model is a linear logistic model in which the intercepts depend on $j$, but the slopes are all equal, i.e.
$$\mathnormal{logit}(\pi_{j}) = \mathnormal{ln}\left(\frac{\pi_{j}}{1 - \pi_{j}}\right) = \alpha_{j} + \beta^{T}x.$$

In other words, proportional odds model takes logistic regression one step further in order to account for ordered categorical responses. For instance, in our analysis we could have used a logistic regression model to investigate the effect of the number of comments on the odds ratio (OR) of "considering $< 3$ topics" vs "considering $\geq 3$ topics". However, in such a case the cut-point would be arbitrary, and we could have used a similar logistic regression model to analyze the effect on the odds ratio (OR) of "considering $< 2$ topics" vs "considering $\geq 2$ topics". In this sense, proportional odds model averages up over all possible cut-point logistic regression models to maximize the amount of information one can get out of the data.

\nolinenumbers

%
%
%
%
%

\bibliography{biblio_final}

\end{document}